\documentclass[%
 reprint,
superscriptaddress,
 amsmath,amssymb,
 aps,
prd,
]{revtex4-2}

\usepackage{graphicx}% Include figure files
\usepackage{dcolumn}% Align table columns on decimal point
\usepackage{bm}% bold math
\usepackage{lipsum}% http://ctan.org/pkg/lipsum
\usepackage{xcolor}
\usepackage[normalem]{ulem}
\usepackage{subfigure}

\usepackage[mathlines]{lineno}% Enable numbering of text and display math
\begin{document}

%\preprint{APS/123-QED}

\title{Low-energy nuclear recoil calibration of the LUX-ZEPLIN experiment with a photoneutron source}% Force line breaks with \\

%\documentclass[superscriptaddress,altaffilletter]{revtex4-1}
%\begin{document}
%\title{LZ Author List for Paper (RevTeX4 Draft 250623T1411(GMT))}
%\date{June 23, 2025 14:11 (GMT)}

% 1 
\author{J.~Aalbers}
\affiliation{SLAC National Accelerator Laboratory, Menlo Park, CA 94025-7015, USA}
\affiliation{Kavli Institute for Particle Astrophysics and Cosmology, Stanford University, Stanford, CA  94305-4085 USA}

% 2 
\author{D.S.~Akerib}
\affiliation{SLAC National Accelerator Laboratory, Menlo Park, CA 94025-7015, USA}
\affiliation{Kavli Institute for Particle Astrophysics and Cosmology, Stanford University, Stanford, CA  94305-4085 USA}

% 3 
\author{A.K.~Al Musalhi}
\affiliation{University College London (UCL), Department of Physics and Astronomy, London WC1E 6BT, UK}

% 4 
\author{F.~Alder}
\affiliation{University College London (UCL), Department of Physics and Astronomy, London WC1E 6BT, UK}

% 5 
\author{C.S.~Amarasinghe}
\affiliation{University of California, Santa Barbara, Department of Physics, Santa Barbara, CA 93106-9530, USA}

% 6 
\author{A.~Ames}
\affiliation{SLAC National Accelerator Laboratory, Menlo Park, CA 94025-7015, USA}
\affiliation{Kavli Institute for Particle Astrophysics and Cosmology, Stanford University, Stanford, CA  94305-4085 USA}

% 7 
\author{T.J.~Anderson}
\affiliation{SLAC National Accelerator Laboratory, Menlo Park, CA 94025-7015, USA}
\affiliation{Kavli Institute for Particle Astrophysics and Cosmology, Stanford University, Stanford, CA  94305-4085 USA}

% 8 
\author{N.~Angelides}
\affiliation{Imperial College London, Physics Department, Blackett Laboratory, London SW7 2AZ, UK}

% 9 
\author{H.M.~Ara\'{u}jo}
\affiliation{Imperial College London, Physics Department, Blackett Laboratory, London SW7 2AZ, UK}

% 10 
\author{J.E.~Armstrong}
\affiliation{University of Maryland, Department of Physics, College Park, MD 20742-4111, USA}

% 11 
% ...

\collaboration{The LUX-ZEPLIN (LZ) Collaboration}
%\maketitle
%\end{document}

\date{\today}% It is always \today, today,

\begin{abstract}
The LZ experiment is a liquid xenon time-projection chamber (TPC) searching for evidence of particle dark matter interactions. In the simplest assumption of elastic scattering, many dark matter models predict an energy spectrum which rises quasi-exponentially with decreasing energy transfer to a target atom. LZ has recently reported evidence for coherent elastic neutrino-nucleus scattering (CE$\nu$NS) of $^{8}$B solar neutrinos, a signal nearly identical to that expected from a dark matter particle with a mass of about 5.5 GeV/$c^{2}$, which similarly produces nuclear recoil energies $<$~5~keV$_{\text{nr}}$. This analysis details the first \textit{in situ} photoneutron calibration of the LZ detector, using a custom yttrium-beryllium source that produces quasi-monoenergetic neutrons. Elastic nuclear recoils from both the dominant 152~keV and the rare 950~keV photoneutron modes are observed at rates consistent with expectation. We report the ionization yield, and the corresponding scintillation yield inferred from it, at a mean detected recoil energy of 3.1 keV$_{\text{nr}}$ and a drift field of 193 V/cm. The measurement demonstrates the ability of LZ to detect nuclear recoils near threshold and provides an independent check on the detector-response framework underlying few-GeV dark-matter and $^{8}$B CE$\nu$NS searches.

\end{abstract}

\maketitle

\section{Introduction}
\label{sec:intro}

Abundant astrophysical evidence suggests the existence of dark matter \cite{doi:10.1126/science.1261381, Clowe:2006eq, BOSS:2013rlg, Planck:2018vyg}, a non-relativistic and non-baryonic matter component of the universe. Despite significant effort, the particle nature of the dark matter remains an open question. Experiments such as LUX-ZEPLIN (LZ) \cite{LZ:2022lsv} seek to address this question by looking for evidence of dark matter scattering with target nuclei. In many weak-scale dark matter models, the simplest elastic scattering interaction results in a recoil energy spectrum which rises quasi-exponentially with decreasing energy transfer to a target atom. The sensitivity of an experiment is thus strongly dependent on its ability to detect the lowest energy nuclear recoil signals. Notably, the spin-independent scattering of a hypothetical 5.5~GeV/$c^{2}$ dark matter particle with a xenon target is calculated to look nearly identical to the coherent elastic neutrino-nucleus scattering (CE$\nu$NS) of $^8$B solar neutrinos, which produce $<$~5~keV nuclear recoils (keV$_{\text{nr}}$)~\cite{Akerib:2022ort}. LZ has since reported evidence for these interactions~\cite{LZ:2025igz}, and expects to accumulate dozens over the course of its campaign~\cite{LZ:2018qzl}. It is therefore critical for the experiment to calibrate the response of recoiling xenon nuclei in this low-energy regime. The calibration reported here is fully independent of  Ref.~\cite{LZ:2025igz}. It was not used as an input, prior, or tuning dataset for that analysis, and it does not enter its likelihood. It was also acquired in a different detector configuration, at a drift field of 193 V/cm rather than 97 V/cm, and is interpreted with an earlier \textsc{NEST} release than the response model of Ref.~\cite{LZ:2025igz}. However, it provides an independent demonstration that LZ detects nuclear recoils in the energy range relevant to those searches, together with a near-threshold yield measurement against which the underlying response-modeling framework can be validated.

Significant effort has been invested in measuring the response of liquid xenon to few-keV nuclear recoils \cite{PhysRevLett.123.231106, LUX:2016ezw, LUX:2022qxb,XENON:2024kbh}. Higher energy neutron sources such as deuterium-tritium (D-T) \cite{PhysRevLett.123.231106}, deuterium-deuterium (D-D) \cite{LUX:2016ezw}, and $^{241}$AmBe ($\alpha$,n) \cite{XENON:2024xgd} can produce low-energy recoils with a small scattering angle; however, these calibrations produce a lower relative fraction of few-keV recoils in liquid xenon. Given the inherent systematic and statistical uncertainty in such low-energy calibrations, it is critical for a discovery experiment to make such calibrations \textit{in situ} with different elastic nuclear recoil spectra such that they are subject to different systematics. 

In this analysis, we address these requirements and report the results of the first photoneutron calibration of the LZ liquid xenon using a custom $^{88}$Y-$^9$Be source \cite{Collar:2013xva, LZ:2024bsz}, colloquially referred to as YBe. The photonuclear reaction on $^{9}$Be proceeds as $^9$Be~+~$\gamma$~$\rightarrow$~$^8$Be~+~n, with a $Q$-value of $-1.66454$~MeV. $^{88}$Y has two measured $\gamma$-rays that can provide this energy: a dominant mode of 1.8361~MeV with 99.2(3)\% branching ratio, and a rare mode of 2.7340 MeV with 0.71(7)\% branching ratio \cite{Robinson:2016kxn}. The emitted photoneutrons are quasi-monoenergetic with kinetic energies of 152~keV and 950~keV with a percent-level angular dependence, $\pm$3~keV and  $\pm$10~keV, respectively. In the case of elastic scattering, the maximum energy transferred by a neutron, of mass $m$, in a single-scatter (SS) with a xenon nucleus of mass $M$, is $4E_n M m / (M+m)^2 = 4.6$~keV$_{\text{nr}}$ (29~keV$_{\text{nr}}$) given the 152~keV (950~keV) neutrons. The dominant photoneutron mode thus provides a well-defined recoil endpoint against which the low-energy response of a detector such as LZ can be calibrated.

This paper describes a calibration of LZ using monoenergetic photoneutrons from a custom-built YBe source. Observation of the rare 950~keV neutron mode is reported in addition to the dominant 152~keV neutron mode. Additional data, taken with the beryllium replaced with natural magnesium to quantify the $^{88}$Y $\gamma$-ray backgrounds \textit{in situ}, are also presented. In Sec.~\ref{sec:experiment}, we describe the design, deployment, and operation of the source, followed by a discussion of the data selection and associated backgrounds in Sec.~\ref{sec:data}. The nuclear recoil light and charge yield results, along with observations about the energy spectra, are presented in Sec.~\ref{sec:results}.

\begin{figure*}[!t]
	\centering
	\includegraphics[width=0.98\textwidth]{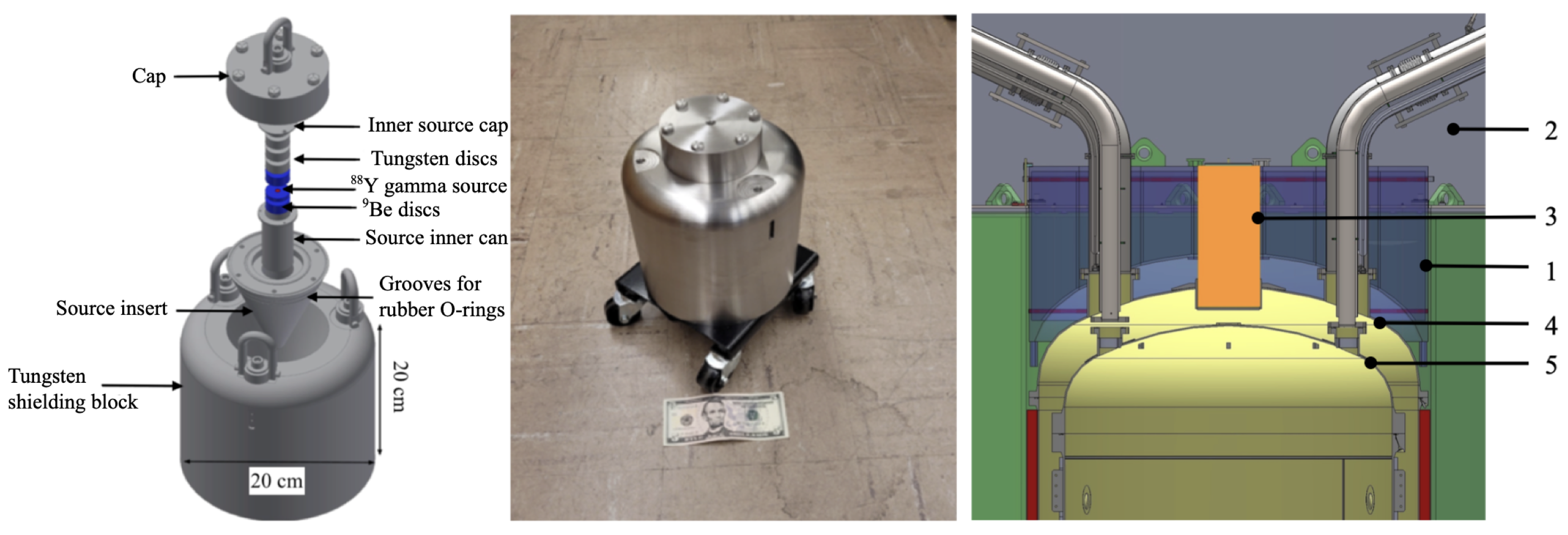}
	\caption{Left: CAD drawing for the YBe photoneutron source assembly. The $^{88}$Y sealed source is sandwiched between $^{9}$Be disks to generate neutrons. Middle: YBe source inside the tungsten shielding placed next to a five-dollar bill for scale. Right: layout of the top part of the outer detector acrylic tanks (1) in both the green and purple volumes and water tank (2). The custom cut-out (3) in the top acrylic tank through which the YBe source is deployed is shown. The outer (4) and inner (5) titanium cryostats are also indicated in the figure. Figure is recreated from Ref.~\cite{LZ:2024bsz}.}
	\label{fig:ybegeom}
\end{figure*}

\section{LZ Detector, Calibration, and Simulation}
\label{sec:experiment}

\subsection{Detector and Calibration Configuration}
\label{sec:experimentA}

The LZ instrument is described in Ref.~\cite{LZ:2019sgr}. Here we reprise a few key details relevant to the present measurement. LZ is a liquid xenon time-projection chamber (TPC) nested within two veto detectors: a liquid xenon ``Skin'' scintillation-only detector and a liquid scintillator Outer Detector (OD). The TPC is instrumented with 494 photomultiplier tubes (PMTs) split into a top and bottom array. For each event, LZ detects a primary scintillation photon signal (S1) and an ionization electron signal (S2). The latter is measured by drifting electrons across the liquid xenon target and then amplifying them via proportional electroluminescence in the vapor above the liquid. By collecting both S1 and S2, the interaction vertex can be reconstructed in three dimensions, with 6 mm precision in $(x,y)$ for S2 areas $>3000$~photons detected (phd)~\cite{LZ:2022ysc}. The TPC and the Skin Detector are housed inside an Inner Cryostat Vessel (ICV) and an Outer Cryostat Vessel (OCV). The OCV is surrounded by acrylic vessels filled with gadolinium-loaded liquid scintillator, as part of the OD. The OCV and OD are housed inside a water tank with the water continuously purified. 

The LZ calibration strategy, including the photoneutron source, is described in detail in Refs.~\cite{LZ:2019sgr,LZ:2024bsz}. The YBe source internals and deployment location within the LZ detector are depicted in Fig.~\ref{fig:ybegeom} as recreated from Ref.~\cite{LZ:2024bsz}. As shown in the left panel, the YBe photoneutron source consists of an $^{88}$Y sealed disk source of 4.7~mm diameter and 4.6~mm height placed between natural beryllium metal disks of 2.54~cm diameter and a total height of 3~cm. A 24~$\mu$m thick nickel layer is plated on the surface of the beryllium metal disks for safe handling. The disks are stacked in a cylinder which is then inserted into a tungsten shield block of 20~cm diameter and 20~cm height. An image of the fully constructed and assembled source is shown in the middle panel of Fig.~\ref{fig:ybegeom}. 

The rightmost panel of Fig.~\ref{fig:ybegeom} depicts how the YBe source is deployed within the LZ experiment and can be compared to other implementations in large-scale liquid xenon TPCs \cite{XENON:2024kbh}. To perform this calibration, the top of the water tank is opened, and the tungsten shield block is lowered into the LZ water tank through an opening in the OD acrylic vessels, until it rests on a purpose-built indent in the top center of the outer cryostat depicted as the orange block in the right panel of Fig.~\ref{fig:ybegeom}. The water tank is then promptly closed to minimize air ingress. During low-background data taking, this indent in the outer cryostat is occupied by a small cylindrical acrylic vessel filled with gadolinium-loaded liquid scintillator (referred to as the ``GdLS plug") to complete the OD coverage. Thus when a calibration is completed, the YBe source and tungsten shielding are replaced with the GdLS plug.

The distance from the $^{88}$Y source to the gate electrode is 86.7~cm. The neutrons must traverse the titanium cryostat vessels, top PMT array, and xenon gas before reaching the active liquid xenon volume. This configuration minimizes the amount of high density material through which the neutrons must pass before reaching liquid xenon in the top of the TPC. The tungsten shield below the source, together with the intervening detector materials, attenuates the $^{88}$Y $\gamma$-rays.

\subsection{Calibration Operations}
\label{sec:experimentB}

Several small but key parameter changes were made to the experimental conditions between the first science data reported in Ref.~\cite{LZ:2022lsv} and the deployment of the photoneutron source. Specifically, both the cathode and gate electrode potentials were moved 2~kV further away from ground, and the anode potential was moved 2.5~kV closer to ground. These changes preserved the nominal 193~V/cm electric field across the liquid xenon target, and decreased the potential difference between the gate and the anode by 0.5~kV. Notably, the drift field is larger than the 97~V/cm field reported in more recent LZ dark matter and $^{8}$B CE$\nu$NS searches \cite{LZ:2024zvo,LZ:2025igz}. The scintillation photon and ionization electron gains, $g_1$ and $g_2$, were characterized in this grid configuration using tritium calibration events in the same spatial selection as used in this analysis. The change in grid configuration decreased both the electroluminescence yield and the electron extraction probability, resulting in an $\sim$25\% decrease in the single-electron size, leading to $g_2=34.9\pm0.6$~phd/electron. The trigger threshold was also adjusted in order to retain the same detection efficiency to few-electron events. The value for $g_1$ remains unchanged from that of the first physics search with $g_1=0.114\pm0.003$~phd/photon. Dedicated spatial corrections were also performed allowing for the definition of the standardized corrected quantities $\text{S}1c$ and $\text{S}2c$ used throughout this analysis.

Following the first science data collection from LZ~\cite{LZ:2022ysc,LZ:2022lsv,LZ:2023poo}, the YBe source was deployed with an initial $^{88}$Y activity of 0.339~MBq with 3\% uncertainty, and 135 hours of calibration data were recorded. The neutron emission rate of the source was determined analytically prior to the calibration. The calculation assumes isotropic $\gamma$-ray emission from the $^{88}$Y source and integrates the probability of photonuclear absorption along each ray path through the beryllium disks, using the source geometry described above, the density of natural beryllium (1.85 g/cm$^{3}$), and the average measured photonuclear cross-sections of 0.651~$\pm$~0.007~mb and 0.592~$\pm$~0.038~mb for the 1.8361~MeV and 2.7340~MeV $\gamma$-rays, respectively, taken from world data in Ref.~\cite{Robinson:2016kxn} and references therein. The resulting neutron-to-gamma ratios, i.e. the probability that a given emitted $\gamma$-ray produces a neutron in the beryllium, are 1 in 9990 and 1 in 10986 for the two $\gamma$-ray energies. Combining these with the source activity during the calibration and the respective branching fractions, the expected neutron rates emitted from the beryllium are 34~$\pm$~1 n/s for the 152~keV neutrons and 0.22~$\pm$~0.03 n/s for the 950~keV neutrons.

About 21 days after these data were recorded, the beryllium metal disks were replaced with magnesium metal disks of the same dimension, the so-called YMg source. The $^{88}$Y source, with activity then at 0.295~MBq (given t$_{1/2} = 106.6$~d), was deployed again to the same location and 24~hours of data were recorded. The $\gamma$-ray attenuation properties of magnesium are within 5\% of those of beryllium, but $\gamma$-rays from $^{88}$Y decays are below the energy threshold for photoneutron production in magnesium. This step, therefore, provided an important ``beam-off'' configuration in which the background from $^{88}$Y decays could be characterized \textit{in situ} without the presence of neutrons. 

\subsection{Simulation of the YBe Source}
\label{sec:experimentC}

The as-built YBe source, configuration, and location were all modeled in Geant4 \cite{Geant4} and included in the same simulation framework as described in Ref.~\cite{LZ:2020zog}. The $^{88}$Y $\gamma$-rays and YBe photoneutrons are simulated as being isotropically emitted from the $^{88}$Y source and beryllium, respectively. These $\gamma$-rays and neutrons are then allowed to propagate through the simulated LZ detector before reaching the active liquid xenon volume where the recoils of interest occur. Particle interactions in the active xenon volumes are then converted into observed quantities, e.g. S1 and S2 signals and reconstructed positions. More details on this detector response are described in Sec.~\ref{sec:results}.

These dedicated Geant4 simulations indicate that the emitted photoneutron energy spectra are degraded by interactions in the intervening materials. These interactions occur between the neutron's production point in the beryllium and its detection in the active liquid xenon. Of the emitted 152 keV and 950~keV neutrons, approximately 1\% and 3\%, respectively, deposit at least 0.5~keV$_{\text{nr}}$ in the TPC. These fractions do not include the effects of the three-fold S1 coincidence, the single-scatter requirement, the fiducial volume, or the analysis selections described in Sec.~\ref{sec:data_sel}. Because the 950~keV recoil spectrum extends to considerably higher energies, where the detection efficiency is larger, the probability that an emitted 950~keV neutron produces an event passing the full selection is approximately nine times that of an emitted 152~keV neutron. The primary sources of energy loss are interactions with the titanium cryostat vessels and the stainless steel and PTFE within the inner cryostat. The resulting energy spectrum of these neutrons entering the TPC falls gradually from the emitted neutron energy toward zero. These same intervening materials are extremely effective at shielding the $^{88}$Y $\gamma$-rays; simulations predict less than 1 in 10$^{4}$ $^{88}$Y decays result in energy deposited in the active liquid xenon. As a result, the neutron-to-gamma ratio for interactions in the TPC is roughly 1 in 100, two orders of magnitude more favorable than the 1 in 9990 ratio at emission, because the intervening material attenuates the $\gamma$-rays far more effectively than the neutrons.

\begin{figure}[!t]
	\centering
	\includegraphics[width=0.98 \columnwidth]{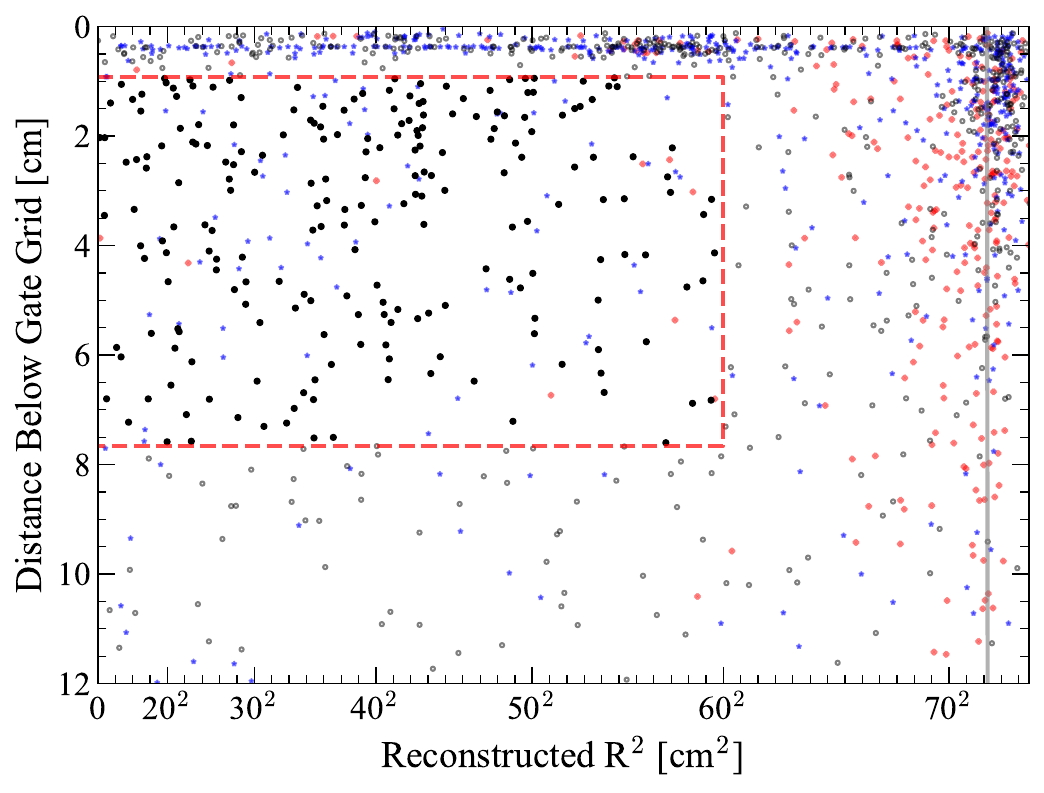}
	\caption{Candidate single-scatter nuclear recoils from the photoneutron source after all data selection criteria (black points) alongside all single-scatter events (gray points) prior to application of the fiducial volume (red-dashed line) defined in Sec.~\ref{sec:data}. Also shown are events removed by the prompt veto (red points) and events that were not delayed-tagged (blue points). The reconstructed wall position is shown as the solid gray line.}
	\label{fig:rz}
\end{figure}

\section{Data Selection and Analysis}
\label{sec:data}

\subsection{Data Selection}
\label{sec:data_sel}

The data selection criteria were largely based on what was used for the first LZ dark matter search \cite{LZ:2022lsv} with a few modifications. In brief, this included a requirement of single-scatter events with a 3-fold PMT coincidence, photon hit patterns consistent with genuine single-scatter particle interactions in the TPC, a trigger hold-off of 20 seconds after passing muon events, and removal of time periods with a full data acquisition buffer. 
 
This analysis is contained within $0<\text{S}1c<40$~phd, with a lower S2 bound of 7 extracted electrons and an upper bound log$_{10}\text{S}2c<4.0$. The lower bound was selected to remove populations of known spurious S2 emission while maintaining an S2 trigger efficiency of 100\%. The S1 bound was selected to include the endpoint of the 950~keV nuclear recoils. We further define the sub-region $0<\text{S}1c<9$~phd and log$_{10}\text{S}2c<3.3$ as the 152~keV neutron mode region of interest (ROI).

The YBe source is centrally located over the TPC. The reconstructed position of single-scatter events is shown in Fig.~\ref{fig:rz}, which represents a small slice at the top of the 145.6~cm tall LZ TPC. Single-scatter events, in which the neutron scatters just once in the active volume, preferentially result in large angle scatters before the neutron exits the detector. Thus, forward-scatters, traveling downward in the active volume, are likely to scatter multiple times. The mean free path for elastic scatters of 152~keV neutrons in liquid xenon is $\sim$14~cm \cite{ENDF}, so the fiducial volume for this result was chosen with drift times between 6~$\mu$s and 50~$\mu$s (137.9~cm~$<$~Z~$<$~144.6~cm), and a reconstructed radial position $R<60$~cm.  The drift bounds maintain high acceptance to genuine single-scatter neutron recoils while rejecting known background populations. The 6~$\mu$s drift bound mitigates the presence of backgrounds originating from the stainless-steel gate electrode, while the 50~$\mu$s drift bound mitigates the presence of backgrounds originating from the top PMT faces which occur in the xenon vapor (see discussion in Sec.~\ref{sec:bkg_gas}). The radial bound mitigates the presence of backgrounds from the PTFE wall where events with small S2 areas can be reconstructed radially inward.

In addition to the standard LZ analyses \cite{LZ:2022lsv,LZ:2024zvo}, several data selection criteria were either developed or refined for this calibration, in order to ensure maximum purity of neutron single-scatters in the data sample: 
\begin{enumerate}
    \item a completely different formulation of the delayed electron/photon \cite{LUX:2020vbj} time hold-off was developed, utilizing the ratio of total event size to largest S2 size. This was necessary because the trigger rate during the calibration was 40~Hz, compared with the $\sim$15~Hz typical during the dark matter search. The original version of this selection, used in other LZ analyses, would have removed a significant amount of this calibration's livetime.  
    \item a stricter selection on the allowed width of S2 signals was implemented. This was required as S2 signals in this dataset are typically narrower than in the usual dark matter search fiducial target, since drifting electrons have less time to diffuse. This helps mitigate accidental coincidence backgrounds (which can have excessive S2 width) as well as gas phase events whose diffusion is different from that in liquid.
    \item a requirement that the fraction of S2 photons detected by the top array of photomultiplier tubes (top-bottom asymmetry or TBA) be consistent with liquid xenon bulk interactions. This selection, stricter than the versions used in previous analyses, provides additional rejection of spurious events produced by scattering in the xenon gas.
    \item a strict criterion on the ``cleanliness" of the single-scatter event window. Events with more than one S2 larger than three electrons following the primary S2 are excluded. This selection targets classification issues related to pile-up of S2 pulses.
\end{enumerate}

The Skin and OD veto detectors help mitigate low-energy Compton scatters from $^{88}$Y $\gamma$-rays and radiogenic backgrounds by requiring no prompt coincidence signals between the TPC S1 and either the Skin or OD within $300$~ns and $500$~ns windows, respectively. The purity of the neutron selection is further enhanced by selecting events in which a delayed coincidence signal ($<1200$~$\mu$s) is present in either the Skin or OD. The definition of prompt and delayed coincidences follows the definition in Ref.~\cite{LZ:2022lsv}. The delayed coincidence neutron tagging efficiency is measured \textit{in situ} from the YBe dataset. Events in the 152~keV mode ROI are selected without the delayed coincidence requirement, and the efficiency is evaluated as the fraction carrying a delayed Skin or OD signal, after subtracting the predicted accidental-coincidence contribution to that sample and correcting for the probability that an untagged event is nonetheless assigned a delayed tag at random. This yields a tagging efficiency of 74~$\pm$~5\%, where the uncertainty is dominated by the statistics of the untagged sample and by the subtraction of the accidental-coincidence contamination. The probability that an event is falsely tagged in a given dataset is a function of the ambient rate in the Skin and OD detectors and the length of the tagging window. This false tagging probability is elevated during the YBe~(YMg) calibrations to 46\%~(37\%) as compared to 5\% during quiescent dark matter search operation. The difference is due to the elevated activity in the OD and Skin due to the high rate $^{88}$Y source. These efficiencies are applied to the calculations described in the next sections.

\begin{figure}[t!]
    \centering
    \includegraphics[width=1 \columnwidth]{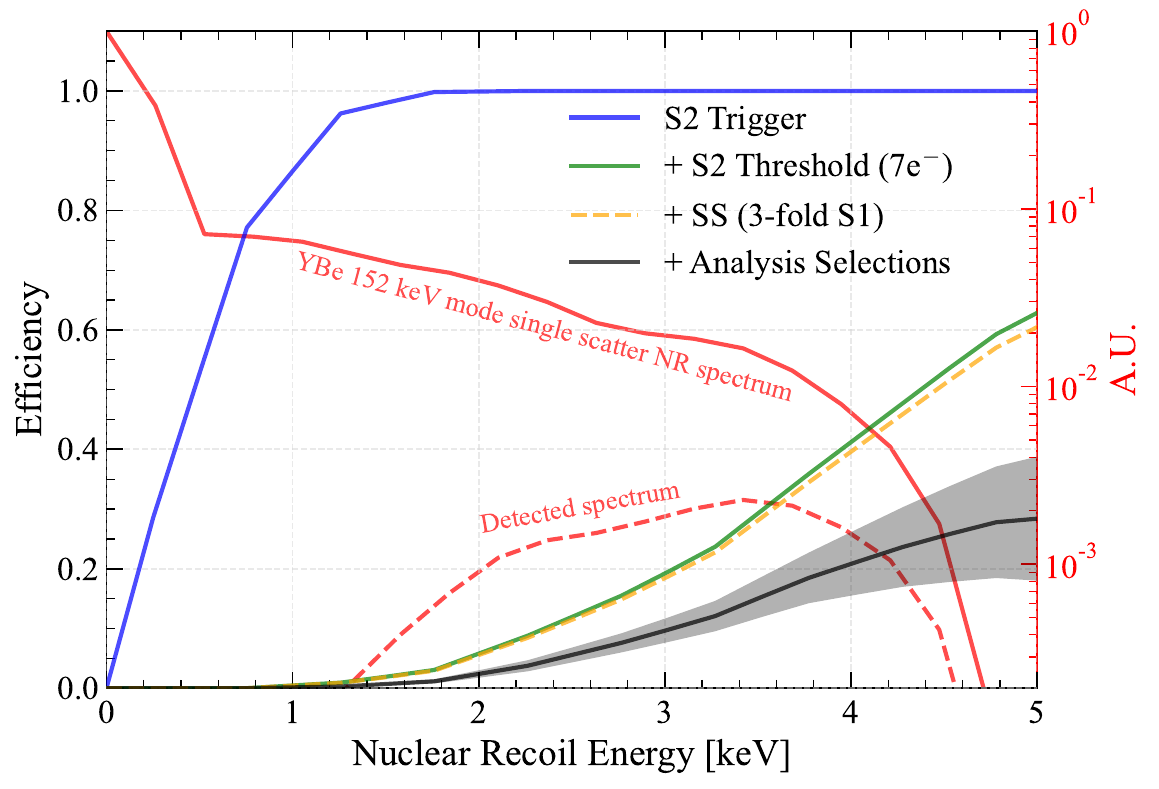}
    \caption{The energy-dependent nuclear recoil signal efficiency is shown after the application, in sequence, of the S2 trigger (blue), S2 threshold (green), 3-fold single-scatter requirement (orange), and analysis selections (black). The shaded gray $\pm$1$\sigma$ uncertainty on the detection efficiency is dominated by the assessments of the analysis selections (primarily the S2 TBA selection). Though not shown, the efficiency continues to increase for higher energies, asymptoting at 33~$\pm$~5\% above 10~keV$_{\text{nr}}$. The solid red line depicts the predicted single-scatter nuclear recoil spectrum of the 152~keV mode YBe photoneutrons. The dashed red line shows the predicted single-scatter recoil spectrum after the application of the final detection efficiency in black. The mean detected YBe recoil energy is 3.1~keV$_{\text{nr}}$. The rightmost axis in red shows the change in the relative rate of the YBe recoil spectra after the application of the detection efficiency.}
    \label{fig:NRDetEff}
\end{figure}

The detection efficiency for low-energy nuclear recoils was evaluated following the procedure in Ref.~\cite{LZ:2023poo} and can be seen in Fig.~\ref{fig:NRDetEff}. The S2 trigger efficiency (blue) was evaluated and determined to be 100\% efficient for S2 signals greater than 4.5~electrons. The subsequent application of the S2 upper and lower bounds is depicted in green. The single-scatter reconstruction (dashed orange) depicts the efficiency to detect $\geq$3-fold S1s and accurately reconstruct these low-energy single-scatters as was done in Ref.~\cite{LZ:2023poo}. The 3-fold, single-scatter and S2 threshold requirements are the dominant limitations on the energy threshold. The final detection efficiency (black) and its $\pm$1$\sigma$ uncertainty (shaded gray) include all of the above efficiencies in addition to the efficiencies associated with the aforementioned selections. Though not shown, the detection efficiency continues to smoothly rise until it asymptotes at 33~$\pm$~5\% efficiency above 10~keV$_{\text{nr}}$. Shown in solid red is the predicted single-scatter recoil spectrum in the active liquid xenon target for the 152~keV mode YBe photoneutrons as predicted from dedicated Geant4 simulations. The rightmost axis in red provides a scale for the relative change in the recoil spectrum induced by the 152~keV mode neutrons after the application of the black detection efficiency. The dashed red line shows the detected 152~keV YBe recoil spectrum after application of all detection efficiencies. The mean energy of the resulting nuclear recoils is 3.1~keV$_{\text{nr}}$.

The efficiency of the analysis selections (see the dashed orange and solid black lines in Fig.~\ref{fig:NRDetEff}) could not be evaluated using external data sources due to the specific physical region of interest and small S2 signals of this calibration. Therefore, the efficiency was estimated in a self-consistent fashion using the data within the YBe 152~keV mode ROI itself. This method is not ideal, since the sample contains a small number of unrejected instrumental backgrounds. We therefore assign a systematic uncertainty to the analysis-selection efficiency. For the 152~(950)~keV photoneutron mode, the uncertainty is estimated to be 24\%~(11\%) and is shown as the shaded gray band in Fig.~\ref{fig:NRDetEff}. This is specifically dominated by the uncertainty on the S2 TBA selection. The uncertainties associated with the S2 trigger efficiency and single-scatter reconstruction are subdominant. The nominal analysis-selection efficiency is the fraction of ROI events passing each selection with that selection removed, and the $\pm1\sigma$ band is the envelope of alternative estimates. Since this estimate does not depend on the yield model, residual correlation with the $Q_y$ fit is considered covered by the propagated $\pm1\sigma$ variation. The uncertainty on the total detection efficiency is factored into the error on the neutron normalizations used in the fitting procedure discussed in Sec.~\ref{sec:results}. The effect of the detection-efficiency uncertainty on both the predicted spectral shape and the normalization is propagated to the best-fit results.

\begin{figure}[ht]
  \centering
  {\includegraphics[width=0.45\textwidth]{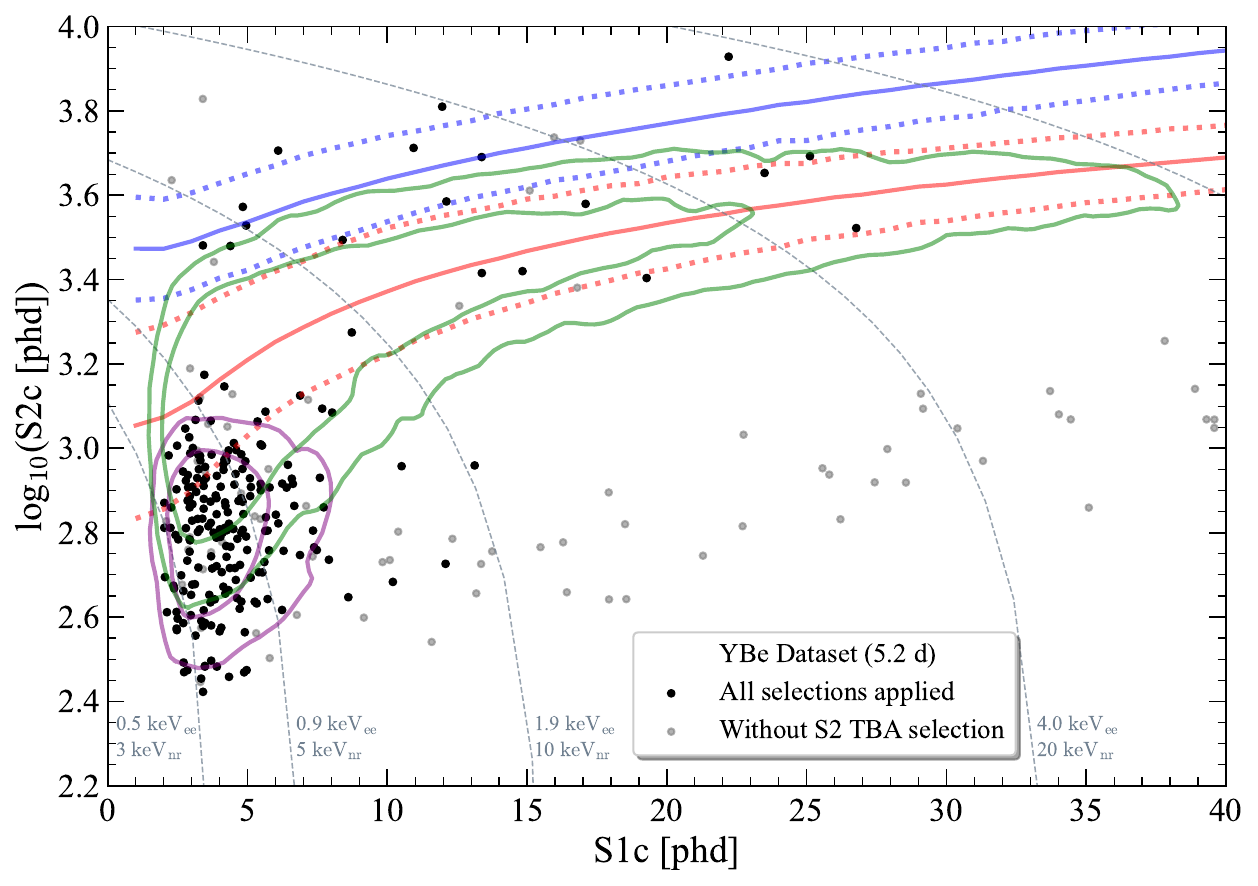}}\\
  {\includegraphics[width=0.45\textwidth]{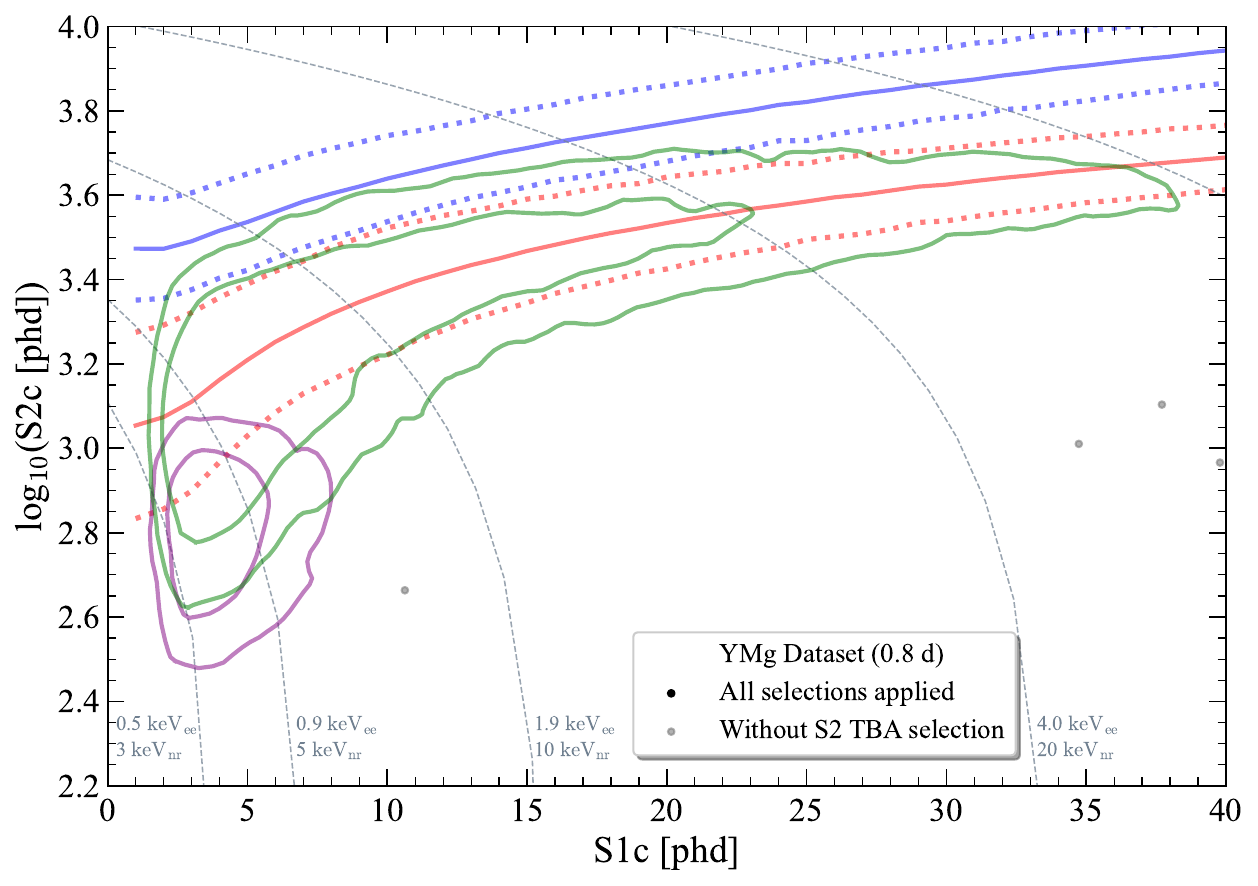}}\\
  {\includegraphics[width=0.45\textwidth]{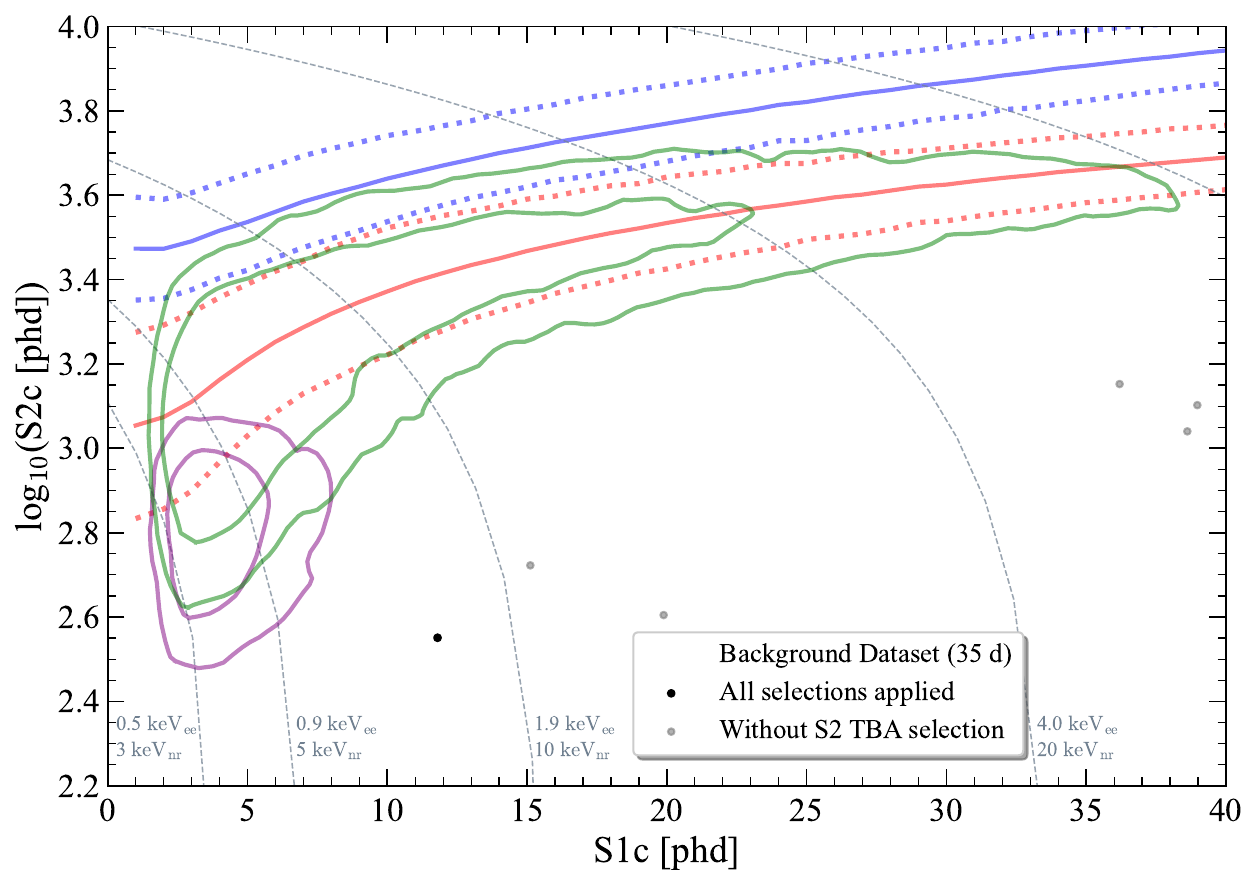}}
  \caption{The YBe photoneutron calibration dataset (top) compared with two control datasets: the YMg calibration (middle) and a background dataset (bottom). Purple contours show the 1$\sigma$ and 2$\sigma$ expectation for 152~keV mode YBe neutron events. Green contours show the 1$\sigma$ and 2$\sigma$ contours for simulated YBe 950~keV mode neutrons. The blue and red bands describe the median and 10\%-90\% quantiles for the expected response to a flat in energy ER and NR spectrum, respectively. Comprehensive details are given in Sec.~\ref{sec:back}.}
  \label{fig:LogS2vsS1}
\end{figure}

\subsection{Backgrounds}\label{sec:back}

These data selections are identically applied to three datasets taken in the same detector configuration, shown in Fig.~\ref{fig:LogS2vsS1}. The top panel of Fig.~\ref{fig:LogS2vsS1} shows the YBe calibration data, corresponding to a 5.2~day exposure. A total of 215 candidate events remain after all selections. The middle panel shows the YMg dataset, corresponding to a 0.8~day exposure, with no candidate neutron events remaining after all selections. The bottom panel shows a 35~day exposure of background data, i.e. without any calibration source present, with only a single event remaining after all selections. The gray points in each plot show events that pass all selections except the S2 TBA criterion, which they fail. In the case of the YBe data, this reveals the presence of an additional band of neutron-induced background events.

Common to all plots in Fig.~\ref{fig:LogS2vsS1} is a series of contours providing context for expected distributions. The response of LZ to flat spectrum electron recoil (ER) and flat-spectrum nuclear recoil (NR) events is given by the blue and red bands. The median and 10\%-90\% quantiles of the respective distributions are shown as solid and dotted lines, respectively. These bands are drawn from the \textsc{NEST}~2.4.0 model~\cite{Szydagis:2021hfh} fit to tritium and D-D calibration data in conjunction with LZ detector response simulation. The expected response of the YBe 152~keV mode photoneutrons was obtained from the same detector response simulation and best-fit \textsc{NEST} model based on the results presented in Sec.~\ref{sec:results}. The 1$\sigma$ and 2$\sigma$ contours of the 152~keV neutron mode elastic nuclear recoils are indicated in purple. The green contours indicate the 1$\sigma$ and 2$\sigma$ boundaries of the 950~keV neutron mode elastic nuclear recoils. The gray dashed lines indicate contours of constant mean energy with the highest contour corresponding to the endpoint of the 950~keV neutron recoil of 29~keV$_{\text{nr}}$.

Of the 215 events remaining in the YBe dataset, shown as black dots in Fig.~\ref{fig:LogS2vsS1}, 193 lie within the 152~keV mode ROI ($0<\text{S}1c<9$~phd and $\log_{10}\text{S}2c<3.3$). This population at low S1$c$, below the NR band, is visually consistent with elastically scattered 152~keV mode photoneutrons, as indicated by the purple contours. Recoils from an at-threshold source are expected to appear below the NR band~\cite{Sorensen:2012ts}, since upward fluctuations in electron-ion recombination are required to produce 3-fold S1s, which in turn reduces the observed S2 size. The clear visibility of the photoneutrons against the $^{88}$Y $\gamma$-rays and radiogenic backgrounds of the YMg and background datasets reflects the low background rate and stable operating conditions during this calibration.

Four further populations are present in the YBe dataset: accidental coincidences, events within the NR band at higher S1 areas, events within the ER band, and events below the NR band extending to higher S1 areas. Each is described in the subsections below and summarized in Table~\ref{tab:Counts}, which gives the predicted and observed counts in the full analysis region and in the ROI. The two regions are quoted separately because the components populate them differently: the 152~keV mode recoils are almost entirely contained within the ROI, whereas the 950~keV mode recoils extend well beyond it to their 29~keV$_{\text{nr}}$ endpoint, and the ER-band events lie outside it entirely. Gas-phase scatters are not modeled and are quoted only in the ROI, where they bear on the yield fit. The best-fit column is obtained in Sec.~\ref{sec:resultsB}.

\subsubsection{Accidental coincidences}
\label{sec:AC}

Accidental coincidences are an instrumental background formed from the random pairing of otherwise unassociated, isolated S1s and S2s within a single event window. This can mimic physical scatters in the active liquid xenon and produce a background underlying the YBe data. Accidental coincidences are a concern in this analysis given the elevated rate of unpaired S1s and S2s, due to the presence of the source. Although the selections defined in the previous section target these topologies, they are imperfect.

Therefore, a model for accidental coincidences is constructed using the same prescription as in Ref.~\cite{LZ:2025igz}. In brief, a high-statistics synthetic dataset is generated from real data by stitching together S1 and S2 pulses from non-scatter events which pass a minimal set of data quality selections. These pulses are paired together at the waveform level as a function of their ambient electron and photon rates, thus capturing the prevailing TPC pulse environment. Unphysical drift time (UDT) events, which have reported drift times exceeding the maximum physical value (i.e. events originating from the cathode), provide a pure population of accidental coincidences. The accidental coincidence model is normalized by scaling the rate of UDT events in the YBe dataset with those in the synthetic dataset. At this stage, we also verify that the data-model agreement gives passing p-values ($>$0.05).

The final accidental coincidence model is constructed by applying the same suite of selections described in Sec.~\ref{sec:data_sel} to the physical drift time synthetic events, scaling their rate based on the UDT-driven normalization. This procedure yields both a model in $\{\text{S}1c,\text{S}2c\}$ and a predicted rate of 10.0~$\pm$~0.9 accidental coincidence events in the full analysis region of the YBe dataset after all selections, of which 8.5~$\pm$~0.8 fall within the 152~keV mode ROI used for the yield fit (Table~\ref{tab:Counts}). The uncertainty on the predicted accidental coincidence rates originates from the number of events at the normalization, the number of unique S1 and S2 pulses in the model passing selections, and the difference in the ratio of physical drift time events failing S1 and/or S2 pulse selections. Critically, the selections defined in Sec.~\ref{sec:data_sel} reject accidental coincidences with high efficiency, leaving them as a subdominant contribution to the observed event count.

\subsubsection{Higher energy nuclear recoils}

The population of events with $9<\text{S}1c<40$ within the NR band is visually consistent with 950~keV mode photoneutron recoils as indicated by the green contours in Fig.~\ref{fig:LogS2vsS1}. Dedicated simulations were performed of the 950~keV mode neutrons emitted from the YBe source to quantify the rate of these events that reach the TPC and pass all selection criteria. Though the predicted emission rate for this rare branching mode is roughly 1/150 that of the dominant mode, driven mainly by the difference in branching fraction, these higher energy neutrons are about nine times more likely to produce an event passing all selections, as discussed in Sec.~\ref{sec:experimentC}. These simulations predict 9.3~$\pm$~1.5 events in the full analysis region, of which 5~$\pm$~1 occur within the $9<\text{S}1c<40$ region, consistent with the observed number. In the 152~keV mode ROI used in the fitting procedure, we predict 4.3~$\pm$~0.7 events (Table~\ref{tab:Counts}). The systematic uncertainty in this prediction is driven by the systematics on the predicted 950~keV neutron rate and estimated neutron detection efficiency. The agreement between the predicted and observed rates provides an \textit{in situ} consistency check of the source-rate calculation and the neutron transport simulation used in this analysis. The lack of NR band events present in the YMg and background datasets further confirms their origin as the 950~keV YBe photoneutrons. Strictly, the 950~keV mode YBe photoneutrons are a background to the 152~keV mode YBe photoneutron signal. However, their typically higher recoil energies and lower rate significantly limit their impact on the fitting results discussed in the next section. Nevertheless, they are included for completeness.

\subsubsection{Electron recoil backgrounds}
\label{sec:bkg_er}

The population of events within the electron recoil (ER) band, as indicated by the blue contours, can originate from several sources: $^{88}$Y $\gamma$-ray Compton scatters, radiogenic backgrounds from $\beta$-decays of internal contaminants, Compton scatters from decays in detector materials, and $\gamma$-ray interactions resulting from neutron captures.

Typical sources of low-energy ER backgrounds include $\beta$-decays of $^{214}$Pb, $^{212}$Pb and $^{85}$Kr; $2\nu\beta\beta$ of $^{136}$Xe; solar neutrino–electron scattering; and Compton scatters from decay products of $^{60}$Co, $^{40}$K, $^{238}$U, and $^{232}$Th present in the detector materials~\cite{LZ:2022ysc}.  We predict $<0.1$ events in both the background and YMg datasets, consistent with Fig.~\ref{fig:LogS2vsS1}, and a rate of 0.16~$\pm$~0.02 events in the YBe dataset.

The number of ER events from $^{88}$Y $\gamma$-rays cannot be directly determined from simulation due to limited statistics. However, we can use the YMg dataset to constrain the rate of $^{88}$Y $\gamma$-ray interactions in the YBe dataset. Using the method of Ref.~\cite{Feldman:1997qc}, we construct a confidence interval and find a $1\sigma$ upper bound of $1.3$ events. By scaling this result by the respective exposures, $^{88}$Y source activity, and false neutron tagging efficiency, we determine the corresponding confidence interval in the YBe dataset to be $<12.1$ events.

In the YBe calibration, electron recoils can also originate from nuclear deexcitation $\gamma$-rays following neutron capture. Geant4 simulations of 152(950)~keV mode neutrons predict neutron captures on titanium in the cryostat, hydrogen or gadolinium present in the outer detector, and the xenon itself~\cite{PhysRevD.106.032007}. These deexcitation $\gamma$-rays can Compton scatter, producing single scatters within the fiducial volume that pass all selection criteria. Based on these simulations, we predict $0.45~\pm~0.02$~($0.08~\pm~0.01$) events from deexcitation $\gamma$-rays in the YBe dataset after all selections.

We construct a $\pm 1\sigma$ confidence interval for the predicted number of events in the ER band of the YBe dataset. In addition to the sources discussed above, we predict $1.2~\pm~0.1$ accidental coincidence events within the ER band. The resulting predicted ER event count lies in the range 2--14 events. The interval is broad and is dominated by the one-sided $^{88}$Y $\gamma$-ray bound obtained from the short, zero-event YMg exposure. The ER band rate is therefore not precisely predicted by this analysis. The 10 events observed in and above the ER band lie within this interval, but we do not claim to have determined their composition. A longer YMg exposure would be required to do so. Importantly, all of these events lie at log$_{10}\text{S}2c > 3.3$ and therefore fall outside the 152 keV mode ROI. They neither interfere with the identification of the 152 keV mode photoneutrons nor enter the fits discussed in the next section. 

\begin{table*}[t]
    \centering
    \caption{Predicted and best-fit event counts for the 5.2~d YBe calibration exposure. The full analysis region is $0<\mathrm{S1}c<40$~phd with S2 larger than 7 extracted electrons and $\log_{10}\mathrm{S2}c<4.0$. The 152~keV mode ROI, used for the yield fit, is the subset with $0<\mathrm{S1}c<9$~phd and $\log_{10}\mathrm{S2}c<3.3$. The second and third columns are pre-fit predictions, derived as described in Sec.~\ref{sec:back}. The fourth column gives the best-fit values of Sec.~\ref{sec:results}. Uncertainties in the fourth column are the statistical uncertainties returned by the fit; the constrained components retain their prior uncertainties. Uncertainties on the totals are summed linearly. The ER entry in the full region is the $\pm1\sigma$ predicted interval of Sec.~\ref{sec:bkg_er}. The ER-like events observed in the YBe dataset lie at $\log_{10}\mathrm{S2}c>3.3$, outside the 152~keV ROI. Gas-phase scatters are not modeled (Sec.~\ref{sec:bkg_gas}) and are quoted in the ROI as the estimated residual contamination with its conservative upper bound in parentheses. Neither the ER nor the gas-phase component enters the fit, and neither is included in the quoted totals. Further details and implications are discussed in the text.}
    \label{tab:Counts}
    \begin{ruledtabular}
    \begin{tabular}{lccc}
    Source & Predicted (full region) & Predicted (152~keV ROI) & Best-fit (152~keV ROI) \\
    \colrule
    YBe 152~keV mode NR     & $148\pm36$   & $146\pm36$  & $179\pm14$ \\
    YBe 950~keV mode NR     & $9.3\pm1.5$  & $4.3\pm0.7$ & $4.6\pm0.7$ \\
    Accidental coincidences & $10.0\pm0.9$ & $8.5\pm0.8$ & $8.6\pm0.8$ \\
    ER backgrounds          & $[2,\,14]$   &  ---       & not fitted \\
    Gas-phase scatters      & ---          & $\sim2$ ($<10$) & not fitted \\
    \colrule
    Total\footnotemark[1]   & $167\pm38$   & $159\pm38$  & $192\pm16$ \\
    \colrule
    Observed                & 215          & \multicolumn{2}{c}{193} \\
    \end{tabular}
    \end{ruledtabular}
    \footnotetext[1]{Excluding the ER and gas-phase components, which are
    not included in the fit.}
\end{table*}

\subsubsection{Scatters in the gas phase}
\label{sec:bkg_gas}

A population of background events below the NR band appears to extend from higher S1 values down into the 152~keV mode photoneutron population. These events are consistent with scatters in the xenon gas above the anode. Such interactions are characterized by their drift times of $\leq$~62~$\mu$s \cite{LZ:2022lsv} and reduced S2 areas resulting from a combination of lower drift field, lower electroluminescence response from electrons drifting toward the anode, and lower light collection efficiency. These events typically form an additional band below the NR band, and are not a direct background during nominal LZ WIMP searches as a result of the characteristic drift time. Many of these events can be excluded by their distinctly larger S2 TBA values. However, full exclusion of these events is limited for small S2 pulses, due to fluctuations in TBA values.

The population of gas events in the YBe dataset originates from a combination of radiogenic interactions (either from $\gamma$-rays from detector materials or beta decays from internal xenon contaminants), $^{88}$Y $\gamma$-rays, and deexcitation $\gamma$-rays following neutron capture.  The background dataset, without any calibration source present, allows for constraining the rate of gas scatters originating from radiogenic interactions. The gas scatters in the YMg dataset, given the rate of gas scatters of radiogenic origin, then allow for constraining the gas scatter event rate from $^{88}$Y $\gamma$-rays. Given these two measurements, the composition of gas scatters in the YBe dataset can be estimated.

One event consistent with a gas scatter remains in the 35~d exposure background dataset following all selections. No events consistent with a gas scatter are observed in the 0.8~d exposure YMg dataset. To enhance the rate of gas interactions to better measure the relative rate of each potential gas population, the S2 TBA criterion is removed in each of the YBe, YMg, and background datasets. This is shown as the band of gray points in Fig.~\ref{fig:LogS2vsS1}. The population of gas events failing the S2 TBA criterion, $\text{S}1c>10$~phd, below the NR band, and with S2 values consistent with the YBe selection ($\text{log}_{10}\text{S2}c<3.3$) are used as a control region to estimate the relative ratios of gas events within the YBe dataset and the gas event rejection efficiency. In this region, 36, 4, and 5 additional events are revealed in the YBe, YMg, and background datasets, respectively.

The relative ratios of the origin of gas interactions within the YBe dataset, estimated using the control region, are found to be $\sim$~2\% from radiogenic scatters, $\sim$~63\% from $^{88}$Y $\gamma$-rays, and $\sim$~35\% from deexcitation $\gamma$-rays. Further, the gas event rejection efficiency is estimated to be 90\% by calculating the combined fraction of events in the control region (45 out of 50) removed in the YBe, YMg, and background datasets. We emphasize that these fractions are indicative only. The $^{88}$Y component is scaled from the four gas-like events observed in the short YMg exposure, and the deexcitation $\gamma$-ray component is obtained as the residual after subtraction, and therefore carries an uncertainty comparable to its own magnitude. It is used here only to establish that $^{88}$Y $\gamma$-rays and neutron-capture $\gamma$-rays contribute to the gas population. It is not transferable to the composition of the ER band population observed in the liquid, which is selected in a different region of $\{\text{S}1c,\text{S}2c\}$.

Unfortunately, the limited exposure of the YMg dataset and lack of robust simulation capability for xenon gas phase response to $\gamma$-rays limit our ability to  build an accurate model for this background. Considering the predicted signal acceptance of the S2 TBA criterion and the estimated gas rejection efficiency, the residual gas-scatter contribution to the 152~keV mode ROI is estimated at approximately 1\% of events passing the final selection, corresponding to roughly two events in the ROI. Since the rejection efficiency is measured in a control region at $\text{S}1c > 10$~phd and the TBA discrimination degrades for the smaller S2 pulses characteristic of the ROI, we conservatively bound the contribution at 5\% of ROI events. The potential impact of including a gas-scatter background component is assessed in Sec.~\ref{sec:resultsB}, and is deemed to be minimal.

\subsubsection{Background summary}
\label{sec:bkg_sum}
Taken together, the three modeled components of Table~\ref{tab:Counts} account for the observed data in both regions. In the 152~keV mode ROI they predict 159~$\pm$~38 events against the 193 observed, and in the full analysis region 167~$\pm$~38 against 215. In both cases the prediction is dominated by the 152~keV mode nuclear recoils and the uncertainty is dominated by the $\pm$24\% uncertainty on the detection efficiency. Of the 22 events lying outside the ROI, 10 are in or above the ER band, consistent with the predicted interval of 2–14 events. Neither the ER backgrounds, all of which lie at log$_{10}\text{S}2c > 3.3$, nor the unmodeled gas-phase scatters, estimated at approximately two events in the ROI with a conservative upper bound of 5\% of ROI events, are included in the quoted totals or in the fit of Sec.~\ref{sec:results}. The impact of the latter on the extracted yield is bounded explicitly in Sec.~\ref{sec:resultsB}.

\section{Light and Charge Yield Analysis}
\label{sec:results}

\subsection{Methodology}
\label{sec:resultsA}

The signal modeling, as described in Ref.~\cite{LZ:2020zog}, utilizes the parametric Noble Element Simulation Technique (\textsc{NEST}) \cite{Szydagis:2021hfh} to convert deposited energy into excitation and ionization signals. The LZ detector response, as validated by LZ calibration data, specifically tritium decays within the sub-volume used in this analysis, further smears this response into observable scintillation and electroluminescence signals. By incorporating the \textsc{NEST} signal generation model and the detector response model, measurements of the nuclear recoil light and charge yields can be performed by fitting the YBe data in the ROI. The NR yield analysis presented here focuses on determining the overall scale of the charge yield and extracting a corresponding light yield.

The \textsc{NEST} NR yield model is described in detail in Ref.~\cite{Szydagis:2022ikv}. In short, \textsc{NEST} provides twelve parameters governing the total number of quanta produced ($N_q$), light yields ($L_y$), and charge yields ($Q_y$), and eight parameters governing the fluctuations of the mean yields. $N_q$ is a simplified version of the Lindhard nuclear recoil quenching model \cite{osti_4153115}, employing a two-parameter power law relation. $Q_{y}$ is described by a term governing the field and density anti-correlation in light and charge yields (i.e. recombination) and a two-parameter sigmoid controlling the roll-off of the yield at sub-keV energies. $L_{y}$ is defined as the difference between $N_q/E$ and $Q_y$ (hence the yields are anti-correlated) with an additional factor allowing for threshold behavior in $L_{y}$ independent of $Q_{y}$. \textsc{NEST} additionally contains parameters to allow for fluctuations in the individual excitations and ionization yields and parameters describing potentially non-binomial recombination fluctuations. 

Mean values and uncertainties for these parameters are provided in Ref.~\cite{Szydagis:2022ikv}; however, their degeneracy in impacting the expected shape and rate of a given recoil distribution must be assessed on a calibration-specific basis. In this analysis, the short exposure of the YBe calibration, and hence sparse statistics in the final dataset, limit the sensitivity to many of these parameters. 

The observed number of S1 photons in the 152~keV mode neutron spectrum is dominated by upward fluctuations. Consequently, the resulting S1 area spectrum is well-approximated by a Poisson distribution with a mean of 3~photons, convolved with the detection efficiency and the average single photoelectron response of the photomultiplier tubes (PMTs). As a result, variations in $L_y$ renormalize the S1 spectrum without altering its shape. As we fix the neutron rate and detection efficiency, the normalization of the 152~keV mode is fixed such that the predicted count in the ROI varies only through its dependence on $L_y$. The total observed neutron rate therefore constrains $L_y$ through the rate alone, and not through any shape information in the $\text{S}1c$ spectrum. This limitation is expected to persist even with a reduced energy threshold that includes 2-fold S1 events. In contrast, the larger number of electrons produced by 152~keV mode elastic NRs offers significant sensitivity to $Q_y$ due to the high collection efficiency of ionization quanta. Unlike $L_y$, the sub-keV roll-off in $Q_y$ is well-constrained by existing data~\cite{LUX:2022qxb,Lenardo:2019fcn}.

\begin{figure*}[t]
  \centering
  \subfigure{\includegraphics[scale=0.48]{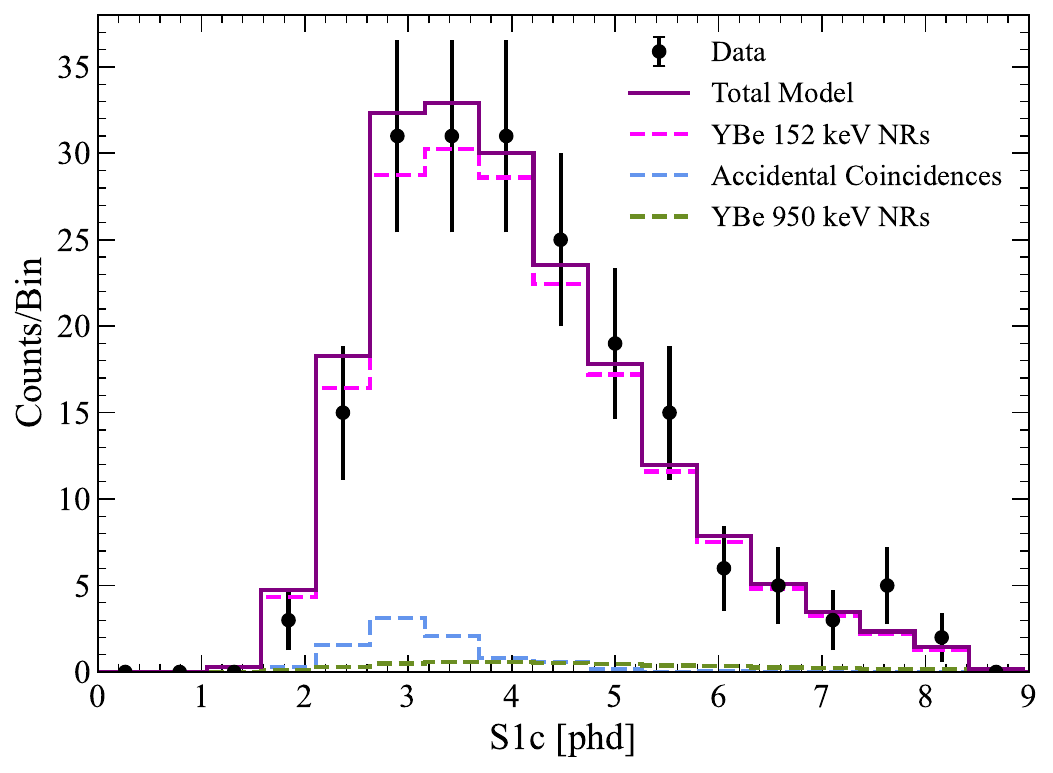}}\quad
  \subfigure{\includegraphics[scale=0.48]{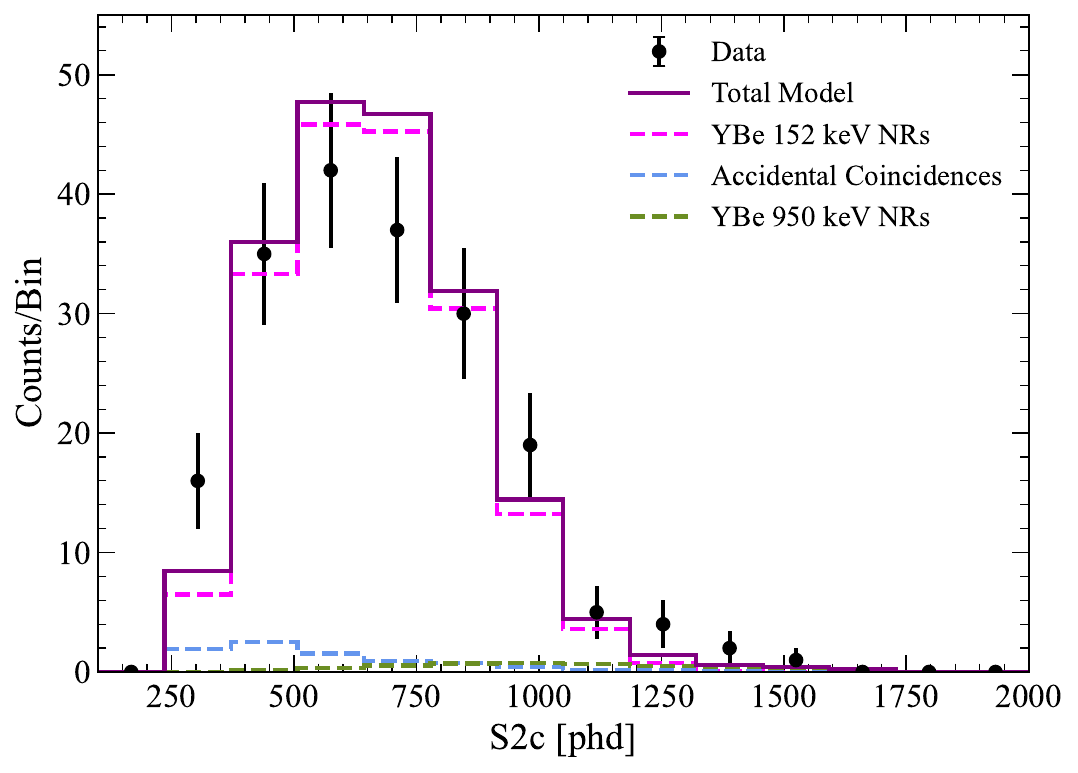}}
  \caption{Projections of the best-fit model onto $\text{S1}c$ (left) and $\text{S2}c$ (right), compared with the YBe data (black points) in the 152~keV mode ROI, $0 < \text{S1}c < 9$~phd and log$_{10}\text{S}2c < 3.3$. The total model (solid purple) comprises the best-fit 152~keV mode YBe elastic nuclear recoils (dashed pink) together with the subdominant accidental-coincidence (dashed blue) and 950~keV mode YBe nuclear recoil (dashed green) contributions. Electron-recoil backgrounds and gas-phase scatters are not modeled and do not enter the fit. The best-fit goodness-of-fit p-values are $0.96$ and $0.13$ for the $\text{S1}c$ and $\text{S2}c$ projections, respectively.}
  \label{fig:fits}
\end{figure*}

Within the YBe 152~keV mode ROI this analysis fits a single parameter, the overall scale of $Q_y$. All other \textsc{NEST} parameters are fixed to their nominal values from Ref.~\cite{Szydagis:2022ikv}. In the energy range relevant here, several \textsc{NEST} parameters act principally to rescale $Q_y$ and are therefore degenerate with one another in shaping the expected recoil distribution. Rather than profile over this degenerate set, we scan the single parameter $\gamma$, which varies $Q_y$ linearly and is thus equivalent to an overall multiplicative scale, and which captures the common scaling behavior while keeping the fit stable and interpretable. This exposure has neither the statistics to break the degeneracy among these parameters nor sensitivity to those governing the sub-keV roll-off of $Q_y$ or the field- and density-dependent recombination terms, the data having been taken at a single drift field and density.

The parameter permitting threshold behavior in $L_y$ independent of $Q_y$ is likewise held fixed, so that $N_q$ is set by the model and $L_y$ follows deterministically from the fitted $Q_y$ via $L_y = N_q/E - Q_y$, where $E$ is the recoil energy. The reported $L_y$ is therefore an inference under the \textsc{NEST} $N_q$ model rather than an independent measurement: the sum $L_y + Q_y = N_q/E$ is fixed by construction and the two are exactly anti-correlated. Both the observed rate, which enters through the $L_y$-dependent acceptance described above, and the shape of the S2$c$ spectrum inform this single parameter. Relaxing the construction by allowing $N_q$ or the independent $L_y$ threshold parameter to vary is not supported by the statistics of this exposure and is deferred to future calibrations.

For each tested $Q_y$ value, the likelihood is formed as a single Poisson term for the total ROI count multiplied by the normalized one-dimensional binned likelihoods of the S1c and S2c spectra; the full two-dimensional S1–S2 correlation is not included. This factorization is appropriate because the $\text{S}1c$ and $\text{S}2c$ spectra in the ROI are largely decoupled. As discussed above, the $\text{S}1c$ spectrum is dominated by upward fluctuations and is largely insensitive to yield variations, constraining primarily the normalization, while the $\text{S}2c$ spectrum carries the $Q_y$ shape information. In addition, the limited statistics of this exposure would not support a stable two-dimensional model.

\subsection{Results}
\label{sec:resultsB}

\begin{figure*}[t]
  \centering
  \subfigure{\includegraphics[scale=0.48]{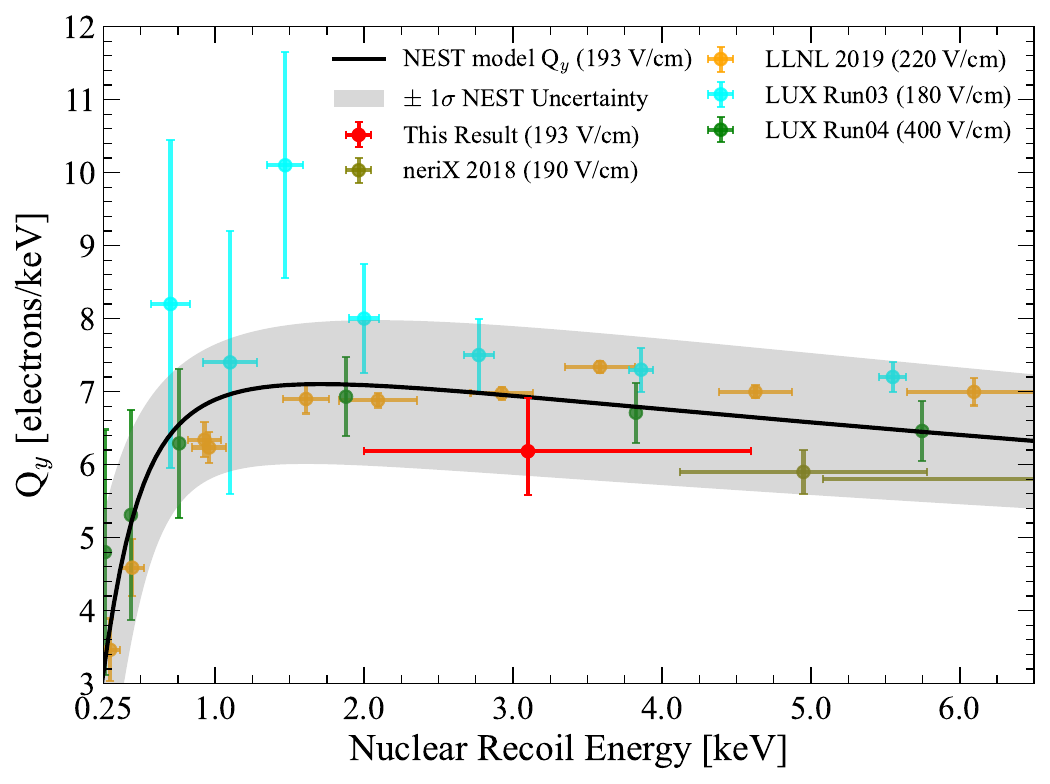}}\quad
  \subfigure{\includegraphics[scale=0.48]{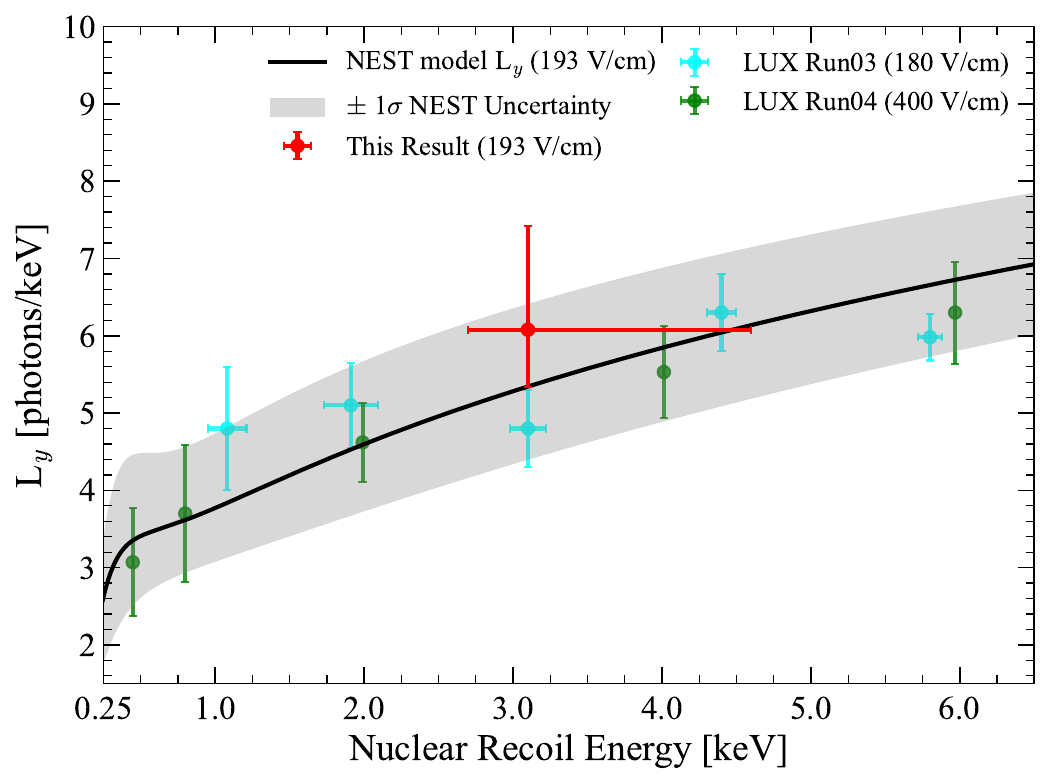}}
  \caption{The charge yield $Q_y$ and light yield $L_y$ as determined from this YBe calibration are shown as red points. The light yield $L_y$ is not measured independently but is inferred through the \textsc{NEST} $N_q$ model via $L_y = N_q/E - Q_y$, which imposes an anti-correlation between $L_y$ and $Q_y$. The two points are therefore exactly anti-correlated and should not be treated as independent constraints. The best-fit $Q_y$ and $L_y$ are quoted at the mean YBe recoil energy of 3.1~keV$_{\text{nr}}$ with a width determined by the minimum recoil energy to which this calibration is sensitive and the 4.6~keV$_{\text{nr}}$ endpoint of the 152~keV photoneutron recoil. The benchmark \textsc{NEST}~2.4.0 model prediction for $Q_y$ and $L_y$ is shown as the black lines along with the $\pm 1\sigma$ uncertainty as a shaded gray band. This result is consistent with the benchmark \textsc{NEST} model and other previous experimental results \cite{LUX:2016ezw,nerix:2018,Lenardo:2019fcn,LUX:2022qxb,huang_thesis}.}
  \label{fig:yields}
\end{figure*}

The pre-fit prediction for the number of 152~keV mode neutrons in the ROI is 146~$\pm$~36 (Table~\ref{tab:Counts}), obtained using the benchmark \textsc{NEST}~2.4.0 model (corresponding to the black line in Fig.~\ref{fig:yields}), the LZ detector response model, the analytically determined neutron rate, and the selections of this analysis. The total uncertainty on the expected neutron counts is the quadrature sum of the errors on the $^{88}$Y activity, $\gamma$-ray branching fraction, photoneutron cross-section, and the systematic on detection efficiency, with the latter being the dominant component. The 950~keV mode YBe photoneutrons were also included in these fits. Given their subdominant prediction of 4.3~$\pm$~0.7 events in the ROI, their spectrum was fixed, though in principle it would be subject to the same variation in yields as the 152~keV mode neutron spectrum. Accidental coincidence backgrounds are also included in the fits, with 8.5~$\pm$~0.8 events (Table~\ref{tab:Counts}) predicted within the ROI.

The best-fit model is shown as the dashed pink line in Fig.~\ref{fig:fits} with the total model shown in solid purple. The YBe data within the ROI are shown as black points.  The best-fit results are summarized in the fourth column of Table~\ref{tab:Counts}. The fitted number of 152~keV mode nuclear recoils, 179~$\pm$~14 (statistical from fit), is consistent with the pre-fit prediction of 146~$\pm$~36, whose $\pm$24\% uncertainty is dominated by the detection efficiency. The 950~keV mode and accidental-coincidence components remain at their predicted values within their constraints. The total of 192~$\pm$~16 events is in good agreement with the 193 events observed in the ROI. The best-fit p-values for the $\text{S}1c$ and $\text{S}2c$ spectra shown in Fig. \ref{fig:fits} are 0.96 and 0.13, respectively, evaluated from the Baker–Cousins likelihood $\chi^{2}$~\cite{BAKER1984437} of each projection. 

The corresponding best-fit charge yield is $0.88^{+0.11}_{-0.09}$ of the nominal \textsc{NEST} value at the mean detected YBe recoil energy of 3.1~keV$_{\text{nr}}$, i.e. +12\%/-10\% of the best-fit value. The normalization uncertainty (24\%) is the quadrature sum enumerated above, comprising the $^{88}$Y activity, the $\gamma$-ray branching fraction, the photoneutron cross-section, and the detection efficiency. This uncertainty on the predicted event rate must be distinguished from its effect on the extracted yield. A $\pm1\sigma$ excursion of the detection efficiency displaces the best-fit $Q_y$ by only +9\%/-8\% (see below), and it is this displacement, not the 24\%, that enters the quoted yield uncertainty. The quoted total is the quadrature sum of this term, the effect of the $g_2$ uncertainty on $Q_y$ (3\%), and the statistical uncertainty returned by the $Q_y$ scan (+7\%/-5\%), the first and last being comparable in size. The $^{88}$Y activity, branching fraction, and photoneutron cross-section uncertainties (3.2\% combined) shift $Q_y$ by $\approx1\%$, and the $g_1$ uncertainty has a negligible impact; both are neglected.

The somewhat low p-value for the $\text{S}2c$ spectrum is in part driven by the second-lowest $\text{S}2c$ bin. An unmodeled gas-phase contribution (Sec.~\ref{sec:bkg_gas}), which is expected to populate low $\text{S}2c$, is a plausible partial contributor to this residual. We do not, however, claim to have positively identified its origin, as the limited YMg exposure and the absence of a validated gas-phase response model preclude a definitive attribution. The central $Q_y$ result is nonetheless largely robust to both this residual and to the modeling of the subdominant accidental-coincidence and gas-phase backgrounds. To bound the maximum possible impact, we repeated the fit after replacing the $\text{S}2c$ template of each of these background populations, in turn, with a single-bin template placed entirely in the second lowest-$\text{S}2c$ bin, which carries the poorest fit residual. The total normalization of each component was held at its predicted value, its $\text{S}1c$ template was left unchanged, and all other components were unaltered; this is the configuration most adversarial to the extracted yield. For the gas-phase component, which has no validated template, a new component was introduced with a normalization equal to the conservative upper bound of 5\% of the events in the ROI and placed entirely in the same bin. In this maximally pessimistic case, the best-fit $Q_y$ shifts upward by less than 3\% for the accidental component and less than 5\% for the gas-phase component. Both shifts are well within the total systematic uncertainty on $Q_y$ and leave the best-fit value within its statistical error. Since no physically plausible mismodeling of the shape of these subdominant components can exceed these bounds, we conclude that the reported $Q_y$ is not compromised by the low $\text{S}2c$ p-value, irrespective of the residual's precise origin.

The effect of the detection-efficiency systematic was assessed by repeating the fit with the $\pm1\sigma$ efficiency curves of Fig.~\ref{fig:NRDetEff}, allowing $Q_y$ to re-minimize in each case. The best-fit $Q_y$ shifts by +9\% under the $+1\sigma$ curve and -8\% under the $-1\sigma$ curve; this displacement is the contribution included in the uncertainty quoted above. The $\text{S}1c$ fit remains very good in both excursions and the $\text{S}2c$ fit remains acceptable under $+1\sigma$, but the $\text{S}2c$ p-value falls below 0.05 under $-1\sigma$, confirming that the systematic acts on the ionization spectrum rather than the scintillation spectrum. We treat the $\pm1\sigma$ curves as a bounding envelope on a systematic that external calibration data cannot constrain, not as alternative response models of comparable likelihood, and read the degradation as an indication that the data disfavor the lower edge of the band. We nonetheless retain the full -8\% displacement in the quoted uncertainty rather than truncating the excursion where the fit remains acceptable, which would have reduced the quoted error.

A key consideration in this analysis is to determine the minimum recoil energy to which the calibration is sensitive. We approached this following the method of Ref.~\cite{LUX:2022qxb}. Using the best-fit model and mean detection efficiencies, the minimum recoil energy considered in the model was set to incrementally increasing cutoff energies; for each energy cutoff, with $Q_y$ held fixed at its best-fit value, the change in the Baker–Cousins $\chi^2$~\cite{BAKER1984437} was calculated. The minimum $L_y$ and $Q_y$ sensitivities are defined as the cutoff at which $\Delta\chi^{2} = 1$, occurring at 2.7~keV$_{\text{nr}}$ for $L_y$ and 2.0~keV$_{\text{nr}}$ for $Q_y$. The higher threshold in the scintillation channel reflects the insensitivity of the $\text{S}1c$ shape to yield variations at the lowest recoil energies as described in Sec.~\ref{sec:resultsA}. To place these thresholds in context, 15\% of expected detected $^{8}$B CE$\nu$NS events, using the recoil spectrum and efficiency of Ref.~\cite{LZ:2025igz}, arise from nuclear recoils above 2.0~keV$_{\text{nr}}$, the regime in which this calibration constrains $Q_y$. The remaining 85\% of the expected $^{8}$B CE$\nu$NS population lies below the energies probed here, so this calibration constrains the response only in the higher energy portion of the detected recoil distribution.

The resulting observed $Q_y$ and $L_y$ are shown as the red points in Fig.~\ref{fig:yields} with respect to the mean (black) and 1$\sigma$ uncertainty (shaded gray) from the \textsc{NEST} model. The best-fit $Q_y$ is a factor of $0.88^{+0.11}_{-0.09}$ of the \textsc{NEST} model at the mean 152~keV mode YBe recoil energy of 3.1~keV$_{\text{nr}}$, as shown in Fig.~\ref{fig:yields}. The light yield $L_y$ is not measured independently but is inferred through the \textsc{NEST} $N_q$ model via $L_y = N_q/E - Q_y$, which imposes an anti-correlation between $L_y$ and $Q_y$. Consequently, the fitted $L_y$ point lying above the \textsc{NEST} median while the fitted $Q_y$ point lies below it is a direct consequence of this model construction.

The widths of the error bars on the $Q_y$ and $L_y$ data points span the range from the minimum recoil energy to which the calibration is sensitive up to the 4.6~keV$_{\text{nr}}$ endpoint of the 152~keV mode elastic recoil spectrum. The total uncertainty on $Q_y$ is as described above. For $L_y$, the total uncertainty also includes the uncertainty on the parameters describing $N_q$, which are used to convert $Q_y$ to $L_y$. Additionally, yield measurements from previous experiments at similar drift fields are presented for comparison.

\section{Summary}
\label{sec:summary}

This is the first photoneutron calibration deployed in the LZ detector, providing an \textit{in situ} measurement of the low-energy nuclear recoil response. Elastic nuclear recoils from the YBe 152~keV and 950~keV photoneutron modes were successfully observed at rates consistent with predictions. The dominant 152~keV neutron mode allowed for calibration of the charge yield, $Q_y$, which is in agreement with existing literature and within \textsc{NEST} model uncertainties. The corresponding $L_y$ is inferred from the fitted $Q_y$ through the \textsc{NEST} $N_q$ model rather than measured independently, and is likewise consistent with that model. Additionally, YMg calibrations were performed with the same $^{88}$Y $\gamma$-ray source, allowing the impact of $^{88}$Y $\gamma$-rays within the liquid xenon to be characterized. No neutron-like events were observed in the YMg or background datasets, and the events observed outside the ROI in the YBe dataset are consistent with the expected ER contributions ($^{88}$Y $\gamma$-ray Compton scatters, neutron-capture $\gamma$-rays, radiogenic backgrounds, and accidental coincidences). Together these support the identification of the low-energy population as YBe photoneutron elastic nuclear recoils. 

Future photoneutron calibrations within LZ would benefit from longer exposures of both YBe and YMg calibrations to accumulate higher statistics. This would allow for comprehensive profiling of the background due to scatters in the xenon vapor above the anode, which are presently an unmodeled, albeit subdominant, background component. Given the source rate, LZ geometry, and YBe source configuration, a few weeks for each of the YBe and YMg datasets would be sufficient. Clearly, it would be desirable to reduce the detector energy threshold by including 2-fold S1 signals. This is presently challenging due to accidental coincidence backgrounds. As discussed in Sec. \ref{sec:resultsA}, it is also likely that including 2-fold S1 signals would have a limited effect on probing lower-energy $L_y$. One may also consider including multi-scatter neutron events if these events can be modeled with sufficient accuracy. Nevertheless, the positive identification of YBe photoneutrons provides direct evidence for LZ's ability to observe near-threshold low-energy nuclear recoils and supports the detector-response framework relevant to few-GeV dark-matter and $^{8}$B CE$\nu$NS searches. These demonstrated benefits lend support to the inclusion of a photoneutron source in calibration plans for next-generation liquid xenon TPCs designed to detect few-GeV mass WIMPs.

\emph{Acknowledgments} - The research supporting this work took place in part at the Sanford Underground Research Facility (SURF) in Lead, South Dakota. Funding for this work is supported by the U.S. Department of Energy, Office of Science, Office of High Energy Physics under Contract Numbers DE-AC02-05CH11231, DE-SC0020216, DE-SC0012704, DE-SC0010010, DE-AC02-07CH11359, DE-SC0015910, DE-SC0014223, DE-SC0010813, DE-SC0009999, DE-NA0003180, DE-SC0011702, DE-SC0010072, DE-SC0006605, DE-SC0008475, DE-SC0019193, DE-FG02-10ER46709, UW PRJ82AJ, DE-SC0013542, DE-AC02-76SF00515, DE-SC0018982, DE-SC0019066, DE-SC0015535, DE-SC0019319, DE-SC0025629, DE-SC0024114, DE-AC52-07NA27344, \& DE-SC0012447. This research was also supported by U.S. National Science Foundation (NSF); the UKRI’s Science \& Technology Facilities Council under award numbers ST/W000490/1, ST/W000482/1, ST/W000636/1, ST/W000466/1, ST/W000628/1, ST/W000555/1, ST/W000547/1, ST/W00058X/1, ST/X508263/1, ST/V506862/1, ST/X508561/1, ST/V507040/1, ST/W507787/1, ST/R003181/1, ST/R003181/2,  ST/W507957/1, ST/X005984/1, ST/X006050/1; Portuguese Foundation for Science and Technology (FCT) under award numbers PTDC/FIS-PAR/2831/2020; the Institute for Basic Science, Korea (budget number IBS-R016-D1); the Swiss National Science Foundation (SNSF)  under award number 10001549. This research was supported by the Australian Government through the Australian Research Council Centre of Excellence for Dark Matter Particle Physics under award number CE200100008. We acknowledge additional support from the UK Science \& Technology Facilities Council (STFC) for PhD studentships and the STFC Boulby Underground Laboratory in the U.K., the GridPP~\cite{faulkner2005gridpp,britton2009gridpp} and IRIS Collaborations, in particular at Imperial College London and additional support by the University College London (UCL) Cosmoparticle Initiative, and the University of Zurich. We acknowledge additional support from the Center for the Fundamental Physics of the Universe, Brown University. K.T. Lesko acknowledges the support of Brasenose College and Oxford University. The LZ Collaboration acknowledges the key contributions of Dr. Sidney Cahn, Yale University, in the production of calibration sources. This research used resources of the National Energy Research Scientific Computing Center, a DOE Office of Science User Facility supported by the Office of Science of the U.S. Department of Energy under Contract No. DE-AC02-05CH11231. We gratefully acknowledge support from GitLab through its GitLab for Education Program. The University of Edinburgh is a charitable body, registered in Scotland, with the registration number SC005336. The assistance of SURF and its personnel in providing physical access and general logistical and technical support is acknowledged. We acknowledge the South Dakota Governor's office, the South Dakota Community Foundation, the South Dakota State University Foundation, and the University of South Dakota Foundation for use of xenon. We also acknowledge the University of Alabama for providing xenon. For the purpose of open access, the authors have applied a Creative Commons Attribution (CC BY) license to any Author Accepted Manuscript version arising from this submission. Finally, we respectfully acknowledge that we are on the traditional land of Indigenous American peoples and honor their rich cultural heritage and enduring contributions. Their deep connection to this land and their resilience and wisdom continue to inspire and enrich our community. We commit to learning from and supporting their effort as original stewards of this land and to preserve their cultures and rights for a more inclusive and sustainable future.

\bibliography{lz-ybe}% Produces the bibliography via BibTeX.

\end{document}